\documentclass[preprint2]{aastex}
\usepackage{graphicx}
\usepackage{color}

\title{Opacity Broadening of $^{13}$CO Linewidths and its Effect on the Variance-Sonic Mach Number Relation}

\author{C. Correia$^{1}$, B. Burkhart$^{2}$, A. Lazarian$^{2}$, V. Ossenkopf$^{3}$, J. Stutzki$^{3}$, J. Kainulainen$^{4}$, G. Kowal$^{5}$ and J. R. de Medeiros$^{1}$}

\affil{$^{1}$~Departamento de F\'isica Te\'orica e Experimental, Universidade Federal do Rio Grande do Norte, 59072-970, Brazil; e-mail: \url{caioftc@dfte.ufrn.br}}

\affil{$^{2}$~Astronomy Department, University of Wisconsin, Madison, 475 N. Charter St., WI 53711, USA}
\affil{$^3$ {Physikalisches Institut der Universit\"{a}t zu K\"{o}ln, Z\"{u}lpicher Strasse 77, 50937 K\"{o}ln, Germany}}
\affil{$^4$ {Max-Planck-Institute for Astronomy, K\"{o}nigstuhl 17, 69117, Heidelberg, Germany}}
\affil{$^5$ {Instituto de Astronomia, Geof\'isica e Ci\^encias Atmosf\'ericas, Universidade de S\~ao Paulo, 05508-090, Brazil}}

\newcommand{\ms}{$\mathrm{M_s}$}
\newcommand{\co}{$^{13}$CO}
\newcommand{\sign}{$\sigma_{N/\langle N\rangle}$}

\begin{abstract}
We study how the estimation of the sonic Mach number (\ms) from $^{13}$CO linewidths relates to the actual 3D sonic Mach number. 
For this purpose we analyze MHD simulations which include post-processing to take radiative transfer effects into account. 
As expected, we find very good agreement between the linewidth estimated sonic Mach number and the actual sonic Mach number
of the simulations for optically thin tracers.  However, we find that opacity broadening causes \ms ~to be overestimated 
by a factor of $\approx1.16-1.3$ when calculated from optically thick $^{13}$CO lines.
We also find that there is a dependency on the magnetic field: super-Alfv\'enic turbulence shows increased line broadening as 
compared with sub-Alfv\'enic turbulence for all values of optical depth for supersonic turbulence.  Our results have implications for the  observationally derived sonic
Mach number--density standard deviation ($\sigma_{\rho/<\rho>}$) relationship, $\sigma^2_{\rho/<\rho>}=b^2\mathrm{M_s^2}$, 
and the related column density  standard deviation (\sign) sonic Mach number relationship. In particular, we find that the parameter $b$,
as an indicator of solenoidal vs. compressive driving, will be underestimated as a result of opacity broadening. 
We compare the \sign--\ms~relation derived from synthetic dust extinction maps and $^{13}$CO linewidths with recent observational studies and 
 find that solenoidally driven MHD turbulence simulations have values of \sign ~which are lower than real molecular clouds.  This may be due
 to the influence of self-gravity which should be included in simulations of molecular cloud dynamics.  
\end{abstract}
\keywords{ISM: structure --- magnetohydrodynamics (MHD) --- methods: numerical}

\begin{document}
\maketitle

\section{Introduction}\label{intro}

Supersonic magnetized turbulence is observed in multiple tracers across several different interstellar media (ISM) phases. 
This includes the neutral medium, as traced by HI \citep[see][]{2010ApJ...714.1398C,2010ApJ...708.1204B,2011ApJ...735..129P}, 
the warm ionized medium, as traced by H$\alpha$ \citep[see][]{2008ApJ...686..363H}, electron density fluctuations \citep{1995ApJ...443..209A,2010ApJ...710..853C}, synchrotron polarization \citep{2011Natur.478..214G,2012ApJ...749..145B} and the molecular medium \citep[see][]{1998ApJS..115..241H,2008ApJ...680..428G},
which includes a variety of molecular traces including the often used carbon monoxide (CO) line. 

Turbulence and magnetic fields in these environments are interrelated to a variety of physical process,
including cosmic ray transport \citep{2004ApJ...614..757Y,2011ApJ...728...60B}, magnetic reconnection \citep{1999ApJ...517..700L,2009ApJ...700...63K}, 
and star formation \citep[and ref. therein]{2007ARA&A..45..565M}.  Furthermore it is clear that, in order to understand MHD turbulence and related 
physical mechanisms, one needs to be able to measure basic plasma parameters such as the sonic and Alfv\'enic Mach numbers ($M_s\equiv V_{turb}/c_s$, 
where $c_s$ is the sound speed, and $M_A\equiv V_{turb}/V_{Alfven}$, respectively).

It is clear from simulations, observations, and theoretical works that compressible turbulence, which generates shocks, 
is important for creating filaments and local regions of high density contrast \citep{2007ApJ...666L..69K}. 
Shocks broaden the gas/dust density and column density probability distribution function (PDF) as well as increase the peaks and 
drag out the distribution's tails towards higher gas densities \citep{2009ApJ...693..250B}. Based on this observation,
several authors \citep[e.g.][to name a few]{1994ApJ...423..681V,1997ApJ...474..730P,2010A&A...512A..81F,2012ApJ...755L..19B} 
have developed relationships between the PDF moments, such as the variance or standard deviation of density ($\sigma_{\rho/<\rho>}$), 
and the sonic Mach number, for example

\begin{equation}
\sigma^2_{\rho/<\rho>}=b^2M_s^2\label{sigbm}
\end{equation}

where b is a parameter that depends on the type of turbulence forcing with b=1/3 for 
pure solenoid forcing and b=1 for pure compressive forcing \citep{2008ApJ...688L..79F}.  
Once gas becomes dense enough, collapse can occur via self-gravity and the PDF forms 
power-law tails towards higher density regions \citep{2000ApJ...535..887K, 2012ApJ...750...13C}.  
Additional variants of Equation 1 have also be developed, e.g. including plasma $\beta$ \citep[see][]{2012MNRAS.423.2680M}.    

Eq. \ref{sigbm} is hard to constrain observationally as volume densities and the 3D velocity structure
are not available from observations.  
Important information on turbulent supersonic motions in molecular clouds comes from the observed non-thermal broadening
of the linewidths of different spectral lines, e.g. from carbon monoxide (CO) emission.  However, CO, in particular \co ~is 
often partially or fully optically thick and traces only a limited dynamic range of column densities 
\citep[see][b]{2009Natur.457...63G,2013ApJ...771..122B}.

Fortunately, dust extinction column density maps of infrared dark clouds (IRDCs), including mid infrared (MIR) and near infrared (NIR)
wavelengths can be used to trace a much larger dynamic range of densities in order to probe the PDF \citep[Av=1--25 for NIR and
Av=10--100 for MIR, see][henceforth KT13]{2001A&A...377.1023L,2011A&A...530A..64K,2013A&A...549A..53K}. However, extinction maps carry no dynamical information 
regarding velocities and for this, molecular line profiles are needed to measure the dynamics of clouds, including the sonic Mach number
for a given cloud temperature.

In this paper we investigate the robustness of measuring the sonic Mach number from maps of $^{13}$CO using synthetic
observations derived from 3D MHD turbulence simulations.  We further create synthetic dust maps in order
to compare the PDF standard deviation with the measured sonic Mach number, following the approach in the observational work of KT13.

\section{Numerical Data}\label{numdata}

\begin{deluxetable}{ r c r r l c r r }
\tablecolumns{10}
\tablewidth{0pc}
\tablecaption{
Description of the 12 simulations each with four different density values (given in column one) in order to probe varying optical depth. Columns two-four give cloud parameters for super-Alfv\'enic simulations while columns five-seven give parameters for sub-Alfv\'enic simulations. Those parameters are: the 1D velocity dispersion, estimated sonic Mach number (see sect. \ref{numdata}) and the average optical depth of the cloud given by SimLine3D \citep[see][]{2002A&A...391..295O}. The actual \ms ~as measured from the original MHD simulations with no radiative transfer is shown above each group of density scaling factors.
\label{tab}}
\tabletypesize{\footnotesize}
\tablehead{
\colhead{}&\multicolumn{3}{c}{Super-Alfv\'enic ($M_{A}$\tablenotemark{a}$\simeq 7.0$)}&
\colhead{}&\multicolumn{3}{c}{Sub-Alfv\'enic ($M_{A}$\tablenotemark{a}$\simeq 0.7$)}\\
\cline{2-4}\cline{6-8}\\
\colhead{density}&\colhead{$\sigma^{1D}_{^{13}CO}$}&\colhead{$M_{S}$\tablenotemark{b}}&\colhead{$\tau$}&&
\colhead{$\sigma^{1D}_{^{13}CO}$}&\colhead{$M_{S}$\tablenotemark{b}}&\colhead{$\tau$}\\
\colhead{}&\colhead{[$kms^{-1}$]}&\colhead{}&\colhead{}&
\colhead{}&\colhead{[$kms^{-1}$]}&\colhead{}&\colhead{}}
\startdata
& \multicolumn{3}{c}{ $M_{S}\tablenotemark{c}\simeq$ 26.2} & & \multicolumn{3}{c}{ $M_{S}\tablenotemark{c} \simeq$ 25.3} \\
    $9n$&6.75&34.4& 0.0017 & & 4.21 & 21.5 & 0.0022 \\
  $275n$&6.29&32.1& 0.03 & & 3.85 & 19.6 & 0.092 \\
 $8250n$&6.25&31.9& 1.2 & & 4.77 & 24.4 & 2.5 \\
$82500n$&6.42&32.8& 2.0 & & 5.64 & 28.8 & 29.6 \\
\cline{2-4} \cline{6-8} \\
& \multicolumn{3}{c}{ $M_{S}\tablenotemark{c}\simeq$  9.0} & & \multicolumn{3}{c}{ $M_{S}\tablenotemark{c}\simeq$  7.9} \\
    $9n$&1.83& 9.3&0.0066 & & 1.26 &  6.4 & 0.0003 \\
  $275n$&1.92& 9.8&0.3 & & 1.41 &  7.2 & 0.3 \\
 $8250n$&2.20&11.2&3.0 & & 1.76 &  9.0 & 3.2 \\
$82500n$&2.32&11.8&54.7 & & 1.94 &  9.9 & 74.4 \\
\cline{2-4} \cline{6-8} \\
& \multicolumn{3}{c}{ $M_{S}\tablenotemark{c}\simeq$  7.1} & & \multicolumn{3}{c}{ $M_{S}\tablenotemark{c}\simeq$  6.8} \\
    $9n$&1.34&6.8&0.008 & & 1.00 &  5.1 & 0.0037 \\
  $275n$&1.47&7.5&0.1 & & 1.12 &  5.7 & 0.4 \\
 $8250n$&1.71&8.7&3.0 & & 1.54 &  7.9 & 10.0 \\
$82500n$&1.81&9.2&68.9 & & 1.68 &  8.6 & 96.1 \\
\cline{2-4} \cline{6-8} \\
& \multicolumn{3}{c}{ $M_{S}\tablenotemark{c}\simeq$  4.3} & & \multicolumn{3}{c}{ $M_{S}\tablenotemark{c}\simeq$  4.5} \\
    $9n$&0.84&4.3&0.0067 & & 0.68 &  3.5 & 0.015 \\
  $275n$&0.84&4.3&0.3 & & 0.85 &  4.4 & 0.4 \\
 $8250n$&1.00&5.1&13.0 & & 1.09 &  5.6 & 13.0 \\
$82500n$&1.06&5.4&56.6 & & 1.18 &  6.0 & 128.0 \\
\cline{2-4} \cline{6-8} \\
& \multicolumn{3}{c}{ $M_{S}\tablenotemark{c}\simeq$  3.1} & & \multicolumn{3}{c}{ $M_{S}\tablenotemark{c}\simeq$  3.2} \\
    $9n$&0.49&2.5&0.016 & & 0.57 &  2.9 & 0.0064 \\
  $275n$&0.57&2.9&0.5 & & 0.61 &  3.1 & 0.1 \\
 $8250n$&0.73&3.7&20.0 & & 0.80 &  4.1 & 14.0 \\
$82500n$&0.78&4.0&140.0 & & 0.86 &  4.4 & 20.3 \\
\cline{2-4} \cline{6-8} \\
& \multicolumn{3}{c}{ $M_{S}\tablenotemark{c}\simeq$  0.7} & & \multicolumn{3}{c}{ $M_{S}\tablenotemark{c}\simeq$  0.7} \\
    $9n$&0.16&0.8&0.082 & & 0.16 &  0.8 & 0.094 \\
  $275n$&0.18&0.9&3.8 & & 0.17 &  0.9 & 4.0 \\
 $8250n$&0.24&1.2&71.0 & & 0.21 &  1.1 & 81.0 \\
$82500n$&0.27&1.4&480.0 & & 0.23 &  1.2 & 806.0 \\
\enddata
\tablenotetext{a}{Initial Alfv\'en Mach Number.}
\tablenotetext{b}{Estimated \ms~from the measured linewidths.}
\tablenotetext{c}{Actual \ms~from original MHD simulations.}
\end{deluxetable}

We generate a database of twelve 3D numerical simulations of isothermal compressible (MHD) turbulence with resolution $512^{3}$ (All models are listed in Table \ref{tab}). For models  with $\mathrm{M_s}<20$ we use the MHD code detailed in \citet{2003MNRAS.345..325C} with large scale solenoidally driven turbulence. Our two models with $\mathrm{M_s}>20$ are from the AMUN code (Kowal et al. 2011). The magnetic field consists of the uniform background field and a time-dependent fluctuating field: $\mathrm{\textbf{B}}=\mathrm{\textbf{B}}_{ext}+\mathrm{\textbf{b(t)}}$; $\mathrm{\textbf{b(0)}}=\mathrm{\textbf{0}}$. For more details on the simulations scheme see \citet[henceforth BL12]{2012ApJ...755L..19B} and \citet{2013ApJ...771..122B}.

Our models are divided in two groups corresponding to sub-Alfv\'enic ($B_{ext}=1.0$) and super-Alfv\'enic ($B_{ext}=0.1$) turbulence. For each group we compute five models with different gas pressures falling into subsonic and supersonic regimes with \ms ~ranging from 0.7 to 26 (see Table \ref{tab}). 

We post-process the simulations to include radiative transfer effects from the $^{13}\mathrm{CO}~J=2-1$ line. 
For more information on this code and the assumptions involved see \citet{2002A&A...391..295O} and \citet[b]{2013ApJ...771..122B}. 
Our original simulation set the sonic Mach number using the sound speed.  
We rescale our simulation's velocity field in such a way that each have the same value of temperature 
and retain the original sonic Mach number. When applying radiative transfer post processing we vary 
the density scaling factor by increasing and decreasing it by a factor of 30 from the standard numerical density of $275cm^{-3}$($cm^{-3}=n$ henceforth), which represents a typical value for a giant molecular cloud. 
 We also include a very high density case with $82500n$. Each line of sight (LoS) 
has a spectral emission of \co, with resolution of $0.5kms^{-1}$ and is taken perpendicular to mean magnetic field\footnote{
A parallel LoS to magnetic field were considered in 4 cases to include low and high sonic(Alfv\'enic) Mach number, see section \ref{resul}.}.
The cube size of our clouds is $5pc$; gas temperature is $10K$; CO Abundance $[^{13}\mathrm{CO/H}]=1.5\times10^{-6}$.

\subsection{Analysis of the Synthetic Observations}\label{analysis}
Once we generate the synthetic \co~line profile maps we measure the dispersion of the  velocity profile using a Gaussian fit. 
Using a gas temperature of $T=10K$ we can calculate the sonic Mach number as:

\begin{equation}
M_s=\frac{\sigma^{3D}_{CO}}{c_{s}}=\frac{\sqrt{3}\sigma^{1D}_{CO}}{c_{s}}\label{Ms}
\end{equation}

Where $\sigma^{1D}_{CO}$ is the velocity dispersion in one dimension and $c_{s}$
  is the sound speed:

\begin{equation}
c_{s}=\sqrt{\frac{K_{b}T}{\mu m_{H}}}\label{Cs}
\end{equation}

In addition to the $^{13}$CO simulations, we create synthetic dust maps using the column density maps from our MHD simulations with line-of-sight taken perpendicular to the mean magnetic field. 
We scale our column density mean value of unity to $2\times10^{22}cm^{-2}$ and then take a simple scaling law \citep[see][]{1978ApJ...224..132B} from column density to extinction as:

\begin{equation}
N_H=1.9\times10^{21}cm^{-2}\frac{A_v}{mag}
\end{equation}

\begin{figure} [!ht]
\plotone{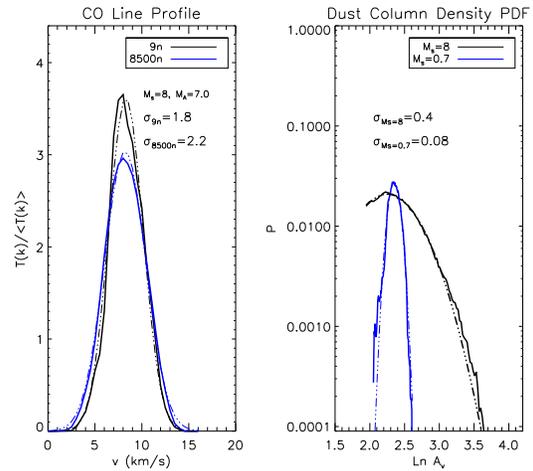}
\figcaption{\footnotesize Left panel: $^{13}$CO line profiles for same supersonic super-Alfv\'enic simulation with different density values.
The solid line is the actual  $^{13}$CO line profile of densities 9n (black) and 
8250n (blue) while thinner dashed line is a 3-term Gaussian fit using the native IDL routine GAUSSFIT. 
Right panel: Synthetic dust column density PDF for simulations with $\mathrm{M_{S}}\approx8$ (black) and $\mathrm{M_{S}}\approx0.7$ (blue); 
Thinner dashed lines is the corresponding Gaussian fit for the PDF. \label{lw-pdf}}
\end{figure}

We then disregarded the column densities with $A_v<7~mag$, following the procedure of KT13.
In the left panel Figure \ref{lw-pdf} we show an example plot of the \co~line profiles for same supersonic super-Alfv\'enic simulation with different density values, hence different
optical depths. 
The opacity broadening effect slightly widens the observed profile.
In the right panel of Figure \ref{lw-pdf} we show an example of the synthetic dust PDFs
(with $\mathrm{A_{V}}$ cut off of 7) of simulations with subsonic(supersonic) Mach number. An increase in sonic Mach number broadens the dust column density PDF distribution.

\placefigure{lw-pdf}

In the following sections, we calculate the observed sonic Mach numbers from $^{13}$CO line maps and compare
this with the actual sonic Mach numbers as measured in the simulations.  We also calculate the PDFs of 
the synthetic dust extinction maps and examine the standard deviation-sonic Mach number relationship as if it
would have been calculated from the observations.

\section{Sonic Mach Number from $^{13}$CO Line Profiles}\label{resul}

\begin{figure} [!ht]
\plotone{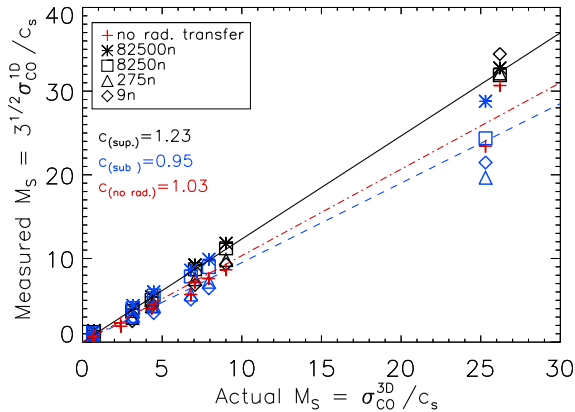}
\figcaption{Measured \ms~from $^{13}\mathrm{CO}~vs.~\mathrm{actual~M_s}$. 
Black solid line is a linear fit to the super-Alfv\'enic clouds; blue dashed 
line is for sub-Alfv\'enic clouds; red dot-dashed for no radiative transfer clouds.\label{ms-ms}}
\end{figure}

Figure \ref{ms-ms} plots  \ms ~as measured from the $^{13}$CO line profiles (measured \ms) vs. actual \ms~ of the simulations.
 We plot four different values of density, which effectively change the optical depth (see Table \ref{tab}). We also over plot linear fits to the
sub-Alfv\'enic (blue lines) and super-Alfv\'enic (black lines) simulations separately with y-intercept held to pass through the origin.
We fit slopes to each optical depth regime and list the values in Table \ref{fits}.

\placefigure{ms-ms}

We find that the optically thin cases (triangle and diamond symbols) reproduce well the actual sonic Mach number from the measured one. However, for our highest optical depth cases the 
ratio between the observationally measured \ms~and the actual \ms~increase to between $\approx1.16-1.3$ (fitted slopes are reported in Table \ref{fits}). 
An additional line broadening due to opacity enlarges the discrepancy between the measured and actual \ms ~as the the optical depth increases.
This occurs because the center of line emission becomes saturated and photons escapes the cloud at the wings of the emission line. Indeed, CO opacity broadening can be explained from the curve-of-growth relationship as well as numerous past studies 
\citep[see Table 2]{1991IAUS..147..451L,1977ApJ...214L..73L,1976ApJ...208..732L}.

Additionally, we observe a magnetic field dependency: super-Alfv\'enic turbulence shows increased line broadening as 
compared with sub-Alfv\'enic turbulence for all values of optical depth for supersonic turbulence.
This effect maybe due to the velocity dispersion of super-Alfvenic turbulence having a
larger dependency on density fluctuations for supersonic turbulence \citep[see][Figures 9,10]{2009ApJ...693..250B}, which causes an additional broadening.
This occurs regardless of the optical depth of the line (see Table \ref{fits}).  We investigated the effect of opacity broadening
on a line-of-sight parallel to the mean magnetic field and found similar results to those reported in Figure 2: the slopes of the measured vs. actual \ms~ covering the
range of optical depth values were 1.06 for sub-Alfv\'enic simulations and 1.28 for super-Alfv\'enic simulations. This indicates that the 
anisotropy present in MHD turbulence is not responsible for the line broadening and instead the strength of the magnetic field is more important.

\section{Standard Deviation-Sonic Mach Number Relation}\label{sigma}

We utilize both our synthetic dust maps and \co ~line dispersion data to estimate the standard deviation-sonic Mach number relation as it would be calculated 
from the observations and compare this with data from KT13. We calculate the standard deviation using two different methods. 
One is direct calculation of the column density standard deviation ($\sigma_{direct}$) divided by the mean value as:

\begin{equation}
\sigma_{N/\langle N\rangle}=\sqrt{\frac{1}{n}\sum^n_{i=1}\left(\frac{N_i}{\langle N\rangle}-\left<\frac{N_i}{\langle N\rangle}\right>\right)^2}\label{sigdir}
\end{equation}

The other method assumes a log-normal distribution and calculates the variance ($\sigma^2_{lnN}$) and mean ($\mu$) values of the logarithm of 
the extinction maps by fitting a Gaussian distribution, illustrated in Figure \ref{lw-pdf} right panel. 
 This method does not assume that we apriori know the actual mean value of the extinction distribution, 
as is indeed the case of the observations, however the assumption of log normality is not always appropriate for all 
molecular clouds as gravity creates deviations from the log-normal distribution \citep[see][]{2012ApJ...750...13C}. 
From the fit we are able to determine the parameters $\mu$ and $\sigma^2_{lnN}$, which are then used to calculate the mean ($\langle N\rangle$) and 
standard deviation ($\sigma_N$) of the corresponding log-normal distribution as:

\begin{equation}
\langle N\rangle=e^{\mu+\sigma^2_{lnN}/2}\label{siglogn1}
\end{equation}
and
\begin{equation}
\sigma_N=[(e^{\sigma^2_{lnN}}-1)e^{2\mu+\sigma^2_{lnN}}]^{1/2}\label{siglogn2}
\end{equation}

For both methods, we calculate the standard deviation for a extinction distribution 
with an $A_\mathrm{v}$ cut off of $\approx7$ for compatibility with the method used in KT13 (hence we denote these as $\sigma_{\mathrm{dir,cut}}$ and $\sigma_{\mathrm{lnN,cut}}$).
 We also investigate the standard deviation calculations for the complete $A_\mathrm{v}$ distribution. This is not available from observations, 
however it is interesting to see how large of a difference in the standard deviation is observed between 
this idealized case and a more realistic distribution with a low $A_\mathrm{v}$ cut off.

\begin{deluxetable}{ r c c }
\tablecolumns{3}
\tablewidth{0pc}
\tablecaption{
Slope (denoted as $c$) for the measured \ms--actual \ms ~data fit. Column two is the slope for a fit including $\mathrm{M_{S}}>20$. Column three is for a fit with $\mathrm{M_{S}}<20$.
\label{fits}}
\tabletypesize{\footnotesize}
\tablehead{\colhead{} & \colhead{$c$ $^a$} & \colhead{$c$ $^b$} }
\startdata
\multicolumn{3}{c}{Super-Alfv\'enic} \\
      $9n$ &  1.25 &  0.99 \\
    $275n$ &  1.19 &  1.06 \\
   $8250n$ &  1.22 &  1.23 \\
  $82500n$ &  1.26 &  1.30 \\
\multicolumn{3}{c}{Sub-Alfv\'enic} \\
      $9n$ &  0.84 &  0.80 \\
    $275n$ &  0.80 &  0.90 \\
   $8250n$ &  1.00 &  1.17 \\
  $82500n$ &  1.16 &  1.28 \\
\enddata
\tablenotetext{a}{Including $\mathrm{M_{S}}>20$.}
\tablenotetext{b}{Only for $\mathrm{M_{S}}<20$}
\end{deluxetable}

We find that there is not a substantial difference between the values of the standard deviation calculated either with 
equation \ref{sigdir} or equations \ref{siglogn1}, \ref{siglogn2}. This is expected for our data since we are dealing 
with log-normal distributions in the case of pure isothermal MHD turbulence. The differences of $\sigma_{\mathrm{{lnN}}}$  
and $\sigma_{\mathrm{direct}}$ extend up to values of 0.15. We additionally find that there is not a large difference between 
the standard deviation of the full column density distribution and the column density distribution which employs a cut off value 
of $A_\mathrm{v}=7$. The differences between $\sigma_{\mathrm{{lnN}}}$ and $\sigma_{\mathrm{direct}}$ with a cut off value 
of $A_\mathrm{v}=7$ extend up to values of 0.13.  However, for a sub-set of real clouds, it may be difficult to discern if one is dealing with 
a true log-normal distribution or beginnings of a high density power-law tail thus the difference in using direct calculation $vs.$ a log-normal 
fit for the calculation of the standard deviation of the column density in observational clouds might be higher.

\begin{figure}[!ht]
\plotone{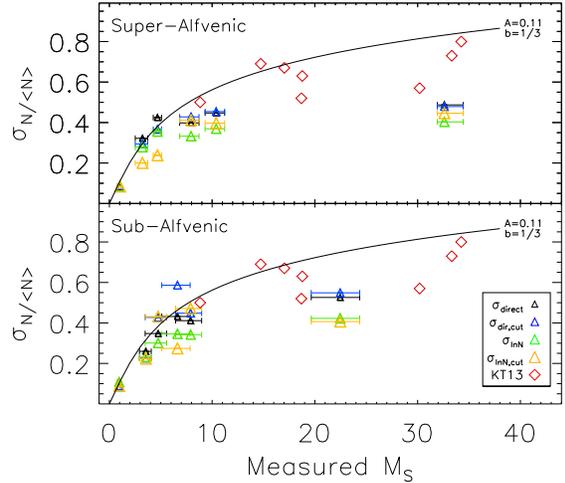}
\figcaption{Column Density dispersion \textit{vs.} measured \ms~for super-Alfv\'enic (upper panel) and sub-Alfv\'enic (lower panel) cases. Clouds from  KT13 table 1 as used in their Figure 7 left frame are shown as red losanges. Black line is a fit to theoretical prediction from BL12.\label{sign-ms-fit}}
\end{figure}

Figure \ref{sign-ms-fit} shows the column density dispersion for the four different methods of
 calculating the standard deviation: $\sigma_{\mathrm{direct}}$, $\sigma_{\mathrm{dir,cut}}$, 
$\sigma_{\mathrm{lnN}}$ and $\sigma_{\mathrm{lnN,cut}}$ (direct calculation, direct calculation 
with $A_\mathrm{v}=7$ cut off, log-normal fit, log-normal fit with $A_\mathrm{v}=7$ cut off, respectively)
  \textit{vs.} estimated \ms ~for sub-Alfv\'enic (top panel) and super-Alfv\'enic cases (bottom panel). 
The values of \ms ~plotted are the average \ms ~calculated from the line widths of the four density cases 
with error bars representing the standard deviation between the values. We perform a linear fit to the [measured \ms, \sign] data, as used in KT13,

\begin{equation}
\sigma_{N/\langle N\rangle}=a_{1}\times M_{s}+a_{2}
\label{sig-msfit}
\end{equation}

Where $a_1$ is the slope between sonic Mach number and column density dispersion and $a_2$ is the intercept.

\placefigure{sign-ms-fit}

Along with the synthetic observations presented in this paper, 
Figure \ref{sign-ms-fit} overplots data from the IRDCs presented in 
KT13 (their Figure 7 left frame -- \ms ~from a Gaussian fit, and reported in KT13 Table 1) 
which are shown with red diamond symbols. The KT13 values of $\sigma_{N/<N>}$ plotted here 
are calculated by method of direct calculation, and thus most compatible with our method of 
direct calculation with Av=7 cut off (blue triangles). Additionally, we include the predicted relation for the column density standard deviation - sonic Mach number
relation from BL12, eq. 4 with $b=1/3$ for solenoidal mixing and $A=0.11$:

\begin{equation}
\sigma_{N/\langle N\rangle}=\sqrt{\left(b^2M_{s}^2+1\right)^A-1}\label{bl12eq}
\end{equation}

The observational points of KT13 consistently show higher values of $\sigma$ than the MHD simulations, 
regardless of the method of calculation for the standard deviation or the Alfv\'en Mach number. 
Upon further inspection, the KT13 points also have a spread in the values of $\sigma$ for a given sonic 
Mach number, suggesting that other physics might come into play in the interpretation of the variance of the data. Processes that could create larger values of variance than expected from solenoidally driven turbulence include gravitational contraction \citep{2012ApJ...750...13C} and compressive forcing \citep{2010A&A...512A..81F,2013A&A...553L...8K}.   

Comparing the mean value of the slopes for the cases of direct calculation of the standard deviation with the A$_v$ cut off, 
$a_1=0.0103\pm0.0047$ ($0.0136\pm0.0050$ for sub-Alfv\'enic; $0.0069\pm0.0044$ for super-Alfv\'enic), with KT13 Figure 7 left frame, which reports $a_1=0.0095\pm0.0066$, we see that there is a good agreement. We see significantly less agreeance of our $a_1$ values from the values of standard deviation derived from the fitted log-normals, 
$a_1=0.0110\pm0.0067$ ($0.0133\pm0.0089$ for sub-Alfv\'enic; $0.0086\pm0.0046$ for super-Alfv\'enic), 
with those of KT13 (their Fig. 8, left frame), $a_1=0.051\pm0.018$. 
The discrepancy maybe due PDF broadening by gravity in the observations (i.e. an unresolved power law tail), thus distorting the standard deviation's relationship to the sonic Mach number.

The other possible explanation for a steeper relation could be that the driving in the KT13 data is compressive forcing, 
however they derived a expression for b,

\begin{equation}
b=\frac{\sigma_{\rho/\langle\rho\rangle}}{M_s}=\frac{\sigma_{N/\langle N\rangle}}{M_s}R^{-1/2}=a_1\times R^{-1/2}\label{b}
\end{equation}

where $R$ is the 3D--2D variance ratio and was found to be between R=[0.03, 0.15] (\citet{2010A&A...513A..67B}). KT13 estimated a b value with 3-$\sigma$ uncertainty of $b=0.20^{+0.37}_{-0.22}$, which indicates that the driving is generally solenoidal to mixed \citep[b=1/3 for solenoidal and b=1 for compressive driving,][]{2008ApJ...688L..79F}.  

However, it is clear that applying a sonic Mach number opacity correction in the case of high optical depths will increase the measured value of b, since the slope $a_1$ will steepen by 
the correction factor. In the case of high optical depth this factor is as high as $\approx1.3$.  Thus the slope, $a_1$, steepens by this factor and, assuming the observations are optically thick,  the value of b calculated by KT13  will increase to $b=0.25^{+0.25}_{-0.15}$. 
These values  still indicate solenoidal to mixed driving. 
Hence the most likely explanation of higher variance values in the KT13 as compared with simulations is the influence of gravity.
This is not unexpected, as \citet[Table 1]{2012ApJ...750...13C} shows that the variance measured from the PDFs in 3D density increases as gravity acts on the cloud.

\section{Discussion}\label{discu}

The models of star formation must invoke stirring by turbulence, including collecting matter by compressible turbulent motions \citep[see][and ref. therein]{2007ARA&A..45..565M} and the removal of magnetic flux from the collapsing region by the process of reconnection diffusion, which is much faster than the traditionally considered process of ambipolar diffusion \citep{2012ApJ...757..154L}. This motivates observational quantitative studies of turbulence and this paper is a part of such studies. 

This paper studies the effect of self-absorption on the observational measurement of the sonic Mach number as measured from \co~linewidths with a range of densities and optical depths. \co~line broadening and extinction PDFs employed in this paper to study the sonic Mach number are in no way the only methods available to researchers to study turbulence in molecular clouds. Sonic\footnote{One can also determine Alfv\'enic Mach number $M_A$ using different contours of isocorrelation obtained with velocity centroids \citep[see][and ref. therein]{2010ApJ...710..125E}, Tsallis statistics \citep{2012ApJ...757..154L,2011ApJ...736...60T}, bispectrum \citep[2013b]{2010ApJ...708.1204B}.} Mach numbers can be also successfully obtained with kurtosis and skewness of the PDF distribution \citep[2010]{2007ApJ...658..423K,2009ApJ...693..250B}. Power spectra of density and velocity can be obtained with Velocity Channel Analysis (VCA) and Velocity Coordinate Spectrum (VCS) techniques \citep[2004, 2006]{1999AAS...195.5102L} which were successfully tested numerically \citep[see][]{2008ApJ...688.1021C} and applied to HI and CO datasets \citep[see][for a review]{2009ApJ...707L.153P,2010ApJ...714.1398C,2009SSRv..143..357L}. In view of complexity of astrophysical turbulence, a simultaneous use of the combination of these techniques presents an unquestionable advantage.

Our empirical finding is that the evaluation of the Mach number from the linewidth is possible even for strongly self-absorbing species, but a correction factor should be applied for opacity broadening.
For the range of absorption depths that we studied we found that this correction factor is at most 1.3. 
It also corresponds to the earlier one dimensional studies of spectral line broadening in the presence of absorption in e.g. \citet{1977ApJ...214L..73L} and \citet{1976ApJ...208..732L}.
An additional consideration is that IRDCs are cold ($T=10-40$K) and are likely to be close to isothermal, but if instead of 10K the cloud had a different temperature there would be a factor of $1/\sqrt{T}$ change in \ms.

\section{Conclusions}\label{conclu}
We create synthetic \co ~emission maps, with varying optical depth,  and dust column density maps from a set of 3D MHD simulations.  
We derive a standard deviation - sonic Mach number relation, as it would be found from the observations, and compare this with recent results from real clouds discussed in KT13.  We find:

\begin{itemize}
\item  Calculations of \ms ~from line widths of \co ~are robust for optically thin \co ~but are overestimated by a factor of up to $\sim1.3$ for optically thick clouds.  This is due to the well known, but often overlooked, effect of opacity broadening.
\item This over estimation of the sonic Mach number as derived from \co ~line widths will cause the slope of the \sign-\ms ~relation to become more shallow and this will result in lower values of the measured b parameter.
\item The \sign~values of clouds reported in KT13 are larger than values found from ideal solenoidally driven simulations of turbulence.  This could be due to the fact that in real molecular clouds there exists the influences of gravity, which will increase the measured column density standard deviation.
\end{itemize}

\acknowledgments 
We thank the anonymous referee for helpful comments.
This work is supported by continuous grants from INEspa\c{c}o/FAPERN/CNPq/MCT.
C.C. acknowledges a graduate PDSE/CAPES grant Process $n^o9392/13-0$.
B.B. acknowledges support from the Wisconsin Space grant.
A.L. acknowledges the Vilas Associate Award and the NSF Grant AST-1212096.
A.L. and B.B. acknowledge the Center for Magnetic Self-Organization in Laboratory and Astrophysical Plasmas and the IIP/UFRN (Natal) for hospitality.
V.O. and J.S. acknowledge support from the Deutsche Forschungsgemeinschaft (DFG) project  $n^o0s~177/2-1$.
V.O., J.S. and J.K. were supported by the central funds of the DFG-priority program 1573 (ISM-SPP).


\begin{thebibliography}{1}
\footnotesize
\bibitem[Armstrong et al.(1995)]{1995ApJ...443..209A} Armstrong, J.~W., 
Rickett, B.~J., \& Spangler, S.~R.\ 1995, \apj, 443, 209 


\bibitem[Beresnyak et al.(2011)]{2011ApJ...728...60B} Beresnyak, A., Yan, 
H., \& Lazarian, A.\ 2011, \apj, 728, 60 


\bibitem[Bohlin et al.(1978)]{1978ApJ...224..132B} Bohlin, R.~C., Savage, B.~D., \& Drake, J.~F.\ 1978, \apj, 224, 132 


\bibitem[Brunt(2010)]{2010A&A...513A..67B} Brunt, C.~M.\ 2010, \aap, 513, A67 


\bibitem[Burkhart et al.(2009)]{2009ApJ...693..250B} Burkhart, B.,  Falceta-Gon{\c c}alves, D., Kowal, G., \& Lazarian, A.\ 2009, \apj, 693, 250 


\bibitem[Burkhart et al.(2010)]{2010ApJ...708.1204B} Burkhart, B., Stanimirovi{\'c}, S., Lazarian, A., \& Kowal, G.\ 2010, \apj, 708, 1204 


\bibitem[Burkhart et al.(2012)]{2012ApJ...749..145B} Burkhart, B., Lazarian, A., \& Gaensler, B.~M.\ 2012, \apj, 749, 145 


\bibitem[Burkhart \& Lazarian(2012)]{2012ApJ...755L..19B} Burkhart, B., \& Lazarian, A.\ 2012, \apjl, 755, L19 (BL12)


\bibitem[Burkhart et al.(2013a)]{2013ApJ...771..123B} Burkhart, B., Lazarian, A., Ossenkopf, V., \& Stutzki, J.\ 2013a, \apj, 771, 123 


\bibitem[Burkhart et al.(2013b)]{2013ApJ...771..122B} Burkhart, B.,Ossenkopf, V., Lazarian, A., \& Stutzki, J.\ 2013b, \apj, 771, 122


\bibitem[Chepurnov et al.(2008)]{2008ApJ...688.1021C} Chepurnov, A., Gordon, J., Lazarian, A., \& Stanimirovic, S.\ 2008, \apj, 688, 1021 


\bibitem[Chepurnov et al.(2010)]{2010ApJ...714.1398C} Chepurnov, A., Lazarian, A., Stanimirovi{\'c}, S., Heiles, C., \& Peek, J.~E.~G.\ 2010, \apj, 714, 1398 


\bibitem[Chepurnov \& Lazarian(2010)]{2010ApJ...710..853C} Chepurnov, A., \& Lazarian, A.\ 2010, \apj, 710, 853 


\bibitem[Cho \& Lazarian(2003)]{2003MNRAS.345..325C} Cho, J., \& Lazarian, A.\ 2003, \mnras, 345, 325 


\bibitem[Collins et al.(2012)]{2012ApJ...750...13C} Collins, D.~C., Kritsuk, A.~G., Padoan, P., et al.\ 2012, \apj, 750, 13 


\bibitem[Esquivel \& Lazarian(2010)]{2010ApJ...710..125E} Esquivel, A., \& Lazarian, A.\ 2010, \apj, 710, 125 


\bibitem[Federrath et al.(2008)]{2008ApJ...688L..79F} Federrath, C., Klessen, R.~S., \& Schmidt, W.\ 2008, \apjl, 688, L79 


\bibitem[Federrath et al.(2010)]{2010A&A...512A..81F} Federrath, C., Roman-Duval, J., Klessen, R.~S., Schmidt, W., \& Mac Low, M.-M.\ 2010, \aap, 512, A81 


\bibitem[Gaensler et al.(2011)]{2011Natur.478..214G} Gaensler, B.~M., 
Haverkorn, M., Burkhart, B., et al.\ 2011, \nat, 478, 214 



\bibitem[Goldsmith et al.(2008)]{2008ApJ...680..428G} Goldsmith, P.~F., 
Heyer, M., Narayanan, G., et al.\ 2008, \apj, 680, 428 


\bibitem[Goodman et al.(2009)]{2009Natur.457...63G} Goodman, A.~A., 
Rosolowsky, E.~W., Borkin, M.~A., et al.\ 2009, \nat, 457, 63 

\bibitem[Heyer et al.(1998)]{1998ApJS..115..241H} Heyer, M.~H., Brunt, C., 
Snell, R.~L., et al.\ 1998, \apjs, 115, 241


\bibitem[Hill et al.(2008)]{2008ApJ...686..363H} Hill, A.~S., Benjamin, 
R.~A., Kowal, G., et al.\ 2008, \apj, 686, 363 


\bibitem[Kainulainen et 
al.(2011)]{2011A&A...530A..64K} Kainulainen, J., Beuther, H., Banerjee, R., Federrath, C., \& Henning, T.\ 2011, \aap, 530, A64 


\bibitem[Kainulainen et 
al.(2013)]{2013A&A...553L...8K} Kainulainen, J., Federrath, C., \& Henning, T.\ 2013, \aap, 553, L8 


\bibitem[Kainulainen 
\& Tan(2013)]{2013A&A...549A..53K} Kainulainen, J., \& Tan, J.~C.\ 2013, \aap, 549, A53 (KT13)


\bibitem[Klessen et al.(2000)]{2000ApJ...535..887K} Klessen, R.~S., Heitsch, F., \& Mac Low, M.-M.\ 2000, \apj, 535, 887 


\bibitem[Kowal et al.(2007)]{2007ApJ...658..423K} Kowal, G., Lazarian, A., 
\& Beresnyak, A.\ 2007, \apj, 658, 423 


\bibitem[Kowal 
\& Lazarian(2007)]{2007ApJ...666L..69K} Kowal, G., \& Lazarian, A.\ 2007, \apjl, 666, L69 


\bibitem[Kowal et al.(2009)]{2009ApJ...700...63K} Kowal, G., Lazarian, A., 
Vishniac, E.~T., \& Otmianowska-Mazur, K.\ 2009, \apj, 700, 63 

\bibitem[Kowal et al.(2011)]{2011NJoP..13}
Kowal, G., Falceta-Gon{\c c}alves, D. A., \& Lazarian, A.,  2011a, New Journal of Physics, 13, 053001

\bibitem[Lazarian 
\& Vishniac(1999)]{1999ApJ...517..700L} Lazarian, A., \& Vishniac, E.~T.\ 1999, \apj, 517, 700 


\bibitem[Lazarian 
\& Pogosyan(1999)]{1999AAS...195.5102L} Lazarian, A., \& Pogosyan, D.\ 1999, Bulletin of the American Astronomical Society, 31, 1449 


\bibitem[Lazarian 
\& Pogosyan(2004)]{2004ApJ...616..943L} Lazarian, A., \& Pogosyan, D.\ 2004, \apj, 616, 943 


\bibitem[Lazarian 
\& Pogosyan(2006)]{2006ApJ...652.1348L} Lazarian, A., \& Pogosyan, D.\ 2006, \apj, 652, 1348 


\bibitem[Lazarian(2009)]{2009SSRv..143..357L} Lazarian, A.\ 2009, \ssr, 
143, 357 


\bibitem[Lazarian et al.(2012)]{2012ApJ...757..154L} Lazarian, A., 
Esquivel, A., \& Crutcher, R.\ 2012, \apj, 757, 154 


\bibitem[Lepine(1991)]{1991IAUS..147..451L} Lepine, J.~R.~D.\ 1991, 
Fragmentation of Molecular Clouds and Star Formation, 147, 451 


\bibitem[Leung \& Liszt(1976)]{1976ApJ...208..732L} Leung, C.-M.,
\& Liszt, H.~S.\ 1976, \apj, 208, 732 


\bibitem[Leung 
\& Brown(1977)]{1977ApJ...214L..73L} Leung, C.~M., \& Brown, R.~L.\ 1977, \apjl, 214, L73 


\bibitem[Lombardi 
\& Alves(2001)]{2001A&A...377.1023L} Lombardi, M., \& Alves, J.\ 2001, \aap, 377, 1023 


\bibitem[McKee 
\& Ostriker(2007)]{2007ARA&A..45..565M} McKee, C.~F., \& Ostriker, E.~C.\ 2007, \araa, 45, 565 


\bibitem[Molina et al.(2012)]{2012MNRAS.423.2680M} Molina, F.~Z., Glover, 
S.~C.~O., Federrath, C., \& Klessen, R.~S.\ 2012, \mnras, 423, 2680 

\bibitem[Ossenkopf(2002)]{2002A&A...391..295O} Ossenkopf, V.\ 2002, \aap, 391, 295


\bibitem[Padoan et al.(1997)]{1997ApJ...474..730P} Padoan, P., Jones, 
B.~J.~T., \& Nordlund, A.~P.\ 1997, \apj, 474, 730 


\bibitem[Padoan et al.(2009)]{2009ApJ...707L.153P} Padoan, P., Juvela, M., 
Kritsuk, A., \& Norman, M.~L.\ 2009, \apjl, 707, L153 


\bibitem[Peek et al.(2011)]{2011ApJ...735..129P} Peek, J.~E.~G., Heiles, 
C., Peek, K.~M.~G., Meyer, D.~M., \& Lauroesch, J.~T.\ 2011, \apj, 735, 129 


\bibitem[Tofflemire et al.(2011)]{2011ApJ...736...60T} Tofflemire, B.~M., 
Burkhart, B., \& Lazarian, A.\ 2011, \apj, 736, 60 


\bibitem[Vazquez-Semadeni(1994)]{1994ApJ...423..681V} Vazquez-Semadeni, E.\ 
1994, \apj, 423, 681 

\bibitem[Yan \& Lazarian(2004)]{2004ApJ...614..757Y} Yan, H., \& Lazarian, A.\ 2004, \apj, 614, 757 

\end{thebibliography}
\end{document}